\documentstyle[amssymb,11pt,epsf]{article}

\newtheorem{theorem}{Theorem}
\newtheorem{lemma}{Lemma}
\newtheorem{definition}{Definition}

\begin{document}

\title{{\bf Mirror Options}}
\author{Juli\'{a}n Manzano\thanks{%
manzano@ecm.ub.es} \\
Departament d'Estructura i Constituents \\
de la Mat\`{e}ria and IFAE,\\
Universitat de Barcelona,\\
Diagonal, 647, E-08028 Barcelona}
\date{}
\maketitle

\begin{abstract}
In this work we present a new family of options (mirror options) specially
crafted to satisfy the necessities of aggressive speculators. The main
ideas behind mirror options are: 1) A product that can be adjusted by the
holder to agree with his/her market view at any time during its life. 2)
The holder's right to make an arbitrary number of those adjustments without
penalizing costs. After defining mirror options as `super-versions' of standard 
options we derive general formulae for their value in the case where the payoff 
is a monotonic function of the underlying (which is the case in calls, puts, 
futures, spreads etc.). We briefly discuss also their valuation for general 
payoffs and the American case. Finally we analyze the situation where 
the number of allowed adjustments is restricted and we point out 
directions for further developments.
\end{abstract}

\vfill
\vbox{
UB-ECM-PF 01/07\null\par
July 2001\null\par
}

\clearpage

\section{Introduction}

Since the first options on stocks appeared in an organized exchange (the
CBOE) in 1973 there has been a continuous growth in worldwide option
markets. This growth is not only apparent in the increasing volumes of
options traded in organized and OTC markets but also in the never ending
production of new products providing alternatives to satisfy the necessities
of investors, banks and other financial institutions \cite{McMillan,Hull}.
These necessities may include risk transference, speculative leveraging,
portfolio diversification, etc.

The purpose of this article is to present a new family of options specially
designed to satisfy the necessities of speculators. Nowadays we have at our
disposal a wide spectrum of products that can virtually serve at whatever
market view that a speculator may have. The main inconvenient with these
`standard' products is that a change in the speculator's market view may
require a change in his/her position forcing unwanted transactions costs.
From that perspective, it would be very convenient for aggressive intra-day
speculators to have a product that could be adjusted to their changing
expectations as many times as desired and without costs associated to these
adjustments. This is exactly the spirit of mirror options.

\subsection{What is a Mirror option?}

Basically a mirror option is a European or American option where the holder
has a well-defined payoff function (for example $\left( S-K\right) _{+}$ if
we are talking about a call) and the right to make an undetermined number of
changes in the real path of the underlying during the life of the option. We
will call these changes `mirrorings' and we will say that the holder has the
right (not the obligation) to `mirror' the underlying an undetermined number
of times. In the next section we will state exactly what we mean by
'mirroring' but now it will be enough to think that the holder has the right
to change the real path of the underlying by a kind of reflected path in the
sense that, after the mirroring, downwards movements in the real path become
upwards movements in the reflected path and vice-versa. The dates for these
mirrorings are not fixed and can be chosen freely by the holder. When the
option is exercised, the payoff is calculated {\em not} with the real
underlying but with the virtual underlying resultant from the mirroring
process performed by the holder up to this time.

The freedom to perform mirrorings whenever during the contract allows the
holder to adapt the option to his/her particular view of the market and,
what is more important, to do it without suffering the associated costs that
this policy would imply trading traditional products like futures, calls,
puts, etc. Taking this into account we think that mirror options offer a new
universe of possibilities that is not available using nowadays standard and
OTC products.

The organization of the article is as follows: in section \ref{defining} we
define the concept of mirroring and immediately after that we state what we
understand by a mirror option in mathematical terms. In section \ref
{valuating} we proceed to valuate mirror options within the simplest market
hypothesis in order to obtain a general valuation formulae. There we obtain
pricing expressions for European mirror options in the case where the payoff
is a monotonic function. We also present as examples the cases of the mirror
call, mirror put and mirror forward. In section \ref{hedging} we show
explicitly a very important feature of the hedging strategy for mirror
options and moved by the conclusions of this section we generalize, in
section \ref{finiteM}, our valuation formulae to the case where there is a
fixed number of allowed mirrorings. Finally in section \ref{conclusions} we
discuss some interesting characteristics of our results and point out
possible alternatives of further work.

\section{Defining mirror options}

\label{defining}
\begin{figure}[ph]
\epsfysize=8.0cm 
\centerline{\epsffile{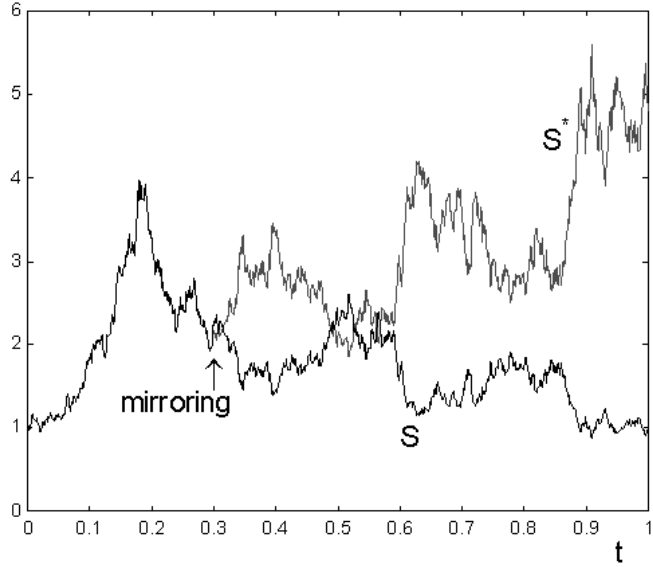}}
\caption{A sample path $S$ (darker line) and its mirror path $S^{\ast }$
(lighter line after mirroring) plotted against time. The volatility $\protect%
\sigma $ and drift $\protect\mu $ for the sample path are taken equal to
one. Aplying It\^{o} lemma we can easily see that the mirror path after the
mirroring has volatility $\protect\sigma ^{\ast }$ and drift $\protect\mu 
^{\ast }$ with $\protect\sigma ^{\ast }=\protect\sigma $ and $\protect\mu 
^{\ast }=\protect\sigma ^{2}-\protect\mu .$}
\label{mirrorfig1}
\end{figure}

\begin{figure}[ph]
\epsfysize=8.0cm 
\centerline{\epsffile{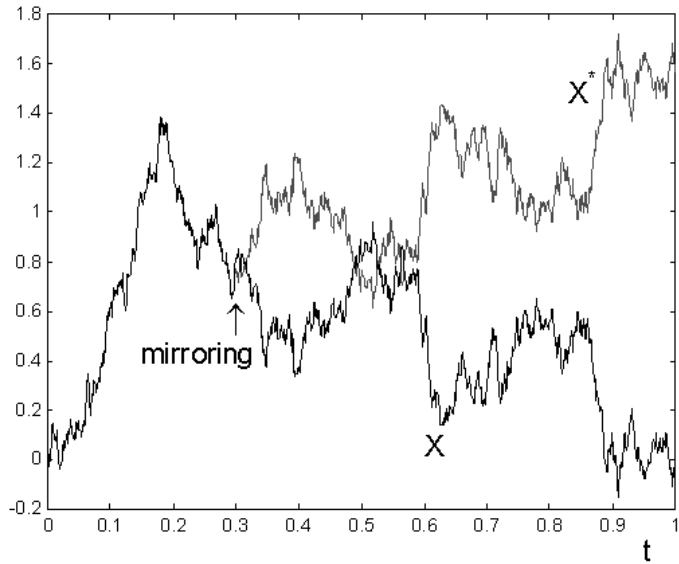}}
\caption{Logarithmic plot of Fig.(\ref{mirrorfig1}) where $X_{t}:=\log
\left( S_{t}/S_{0}\right) $ and $X_{t}^{\ast }:=\log \left( S_{t}^{\ast
}/S_{0}^{\ast }\right) $. Here it is easy to see that after the mirroring $%
X^{\ast }$ becomes the specular reflection of $X$.}
\label{mirrorfig2}
\end{figure}

\begin{figure}[th]
\epsfysize=8.0cm 
\centerline{\epsffile{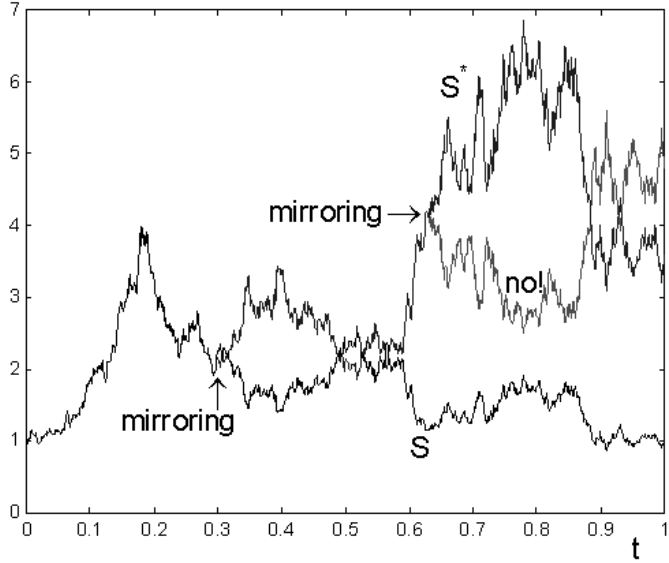}}
\caption{The sample path $S$ of Fig.(\ref{mirrorfig1}) along with the path $%
S^{\ast }$ obtained after two mirrorings. We plot also the mirror path we
would have obtained if the second mirroring were not made, that is, the
mirror path of Fig.(\ref{mirrorfig1}). Note that $S^{\ast }$ after the
second mirroring is just $S$ affected by a multiplicative factor (bigger
than one in this example).}
\label{mirrorfig3}
\end{figure}

In order to give a definition of mirror options we have to
clearly state the meaning of {\em mirroring} and {\em mirror path}. By
mirroring an underlying $S$ at a certain time $t_{m}$ we understand
constructing a new path $S^{\ast }$ (mirror path) defined simply by 
\begin{equation}
S_{t}^{\ast }=\left\{ 
\begin{array}{ccc}
S_{t} &  & t\leq t_{m}, \\ 
S_{t_{m}}^{2}/S_{t} &  & t\geq t_{m},
\end{array}
\right.  \label{single}
\end{equation}
We can see how a mirror path looks like in the example of Fig.(\ref
{mirrorfig1}). In Fig.(\ref{mirrorfig2}) we have plotted the same example in
logarithmic scale where we can clearly see the reason for the name `mirror'
path. In Eq.(\ref{single}) we have given the definition of a mirror path
when a single mirroring time is allowed. If we allow for a second mirroring
at a time $t_{m}^{\prime }$ with $t_{m}^{\prime }\geq t_{m}$ we can define a
new path by just applying Eq.(\ref{single}) over the former single mirrored
path. Note that in the case $t_{m}^{\prime }=t_{m}$ we recover the original
path without mirrorings\footnote{%
we have an involution in mathematical language.}. In general if we allow for
several consecutive mirrorings at times $t_{1}\leq t_{2}\leq \cdots \leq
t_{M}$ it is easy to see that 
\begin{equation}
S_{t}^{\ast }=\left\{ 
\begin{array}{ccc}
S_{t_{1}}^{2}S_{t_{2}}^{-2}\cdots S_{t_{2n}}^{-2}S_{t} &  & t_{0}\leq
t_{1}\leq \cdots \leq t_{2n}\leq t, \\ 
S_{t_{1}}^{2}S_{t_{2}}^{-2}\cdots S_{t_{2n+1}}^{2}S_{t}^{-1} &  & t_{0}\leq
t_{1}\leq \cdots \leq t_{2n+1}\leq t,
\end{array}
\right. \left( n\in {\Bbb N}\right) ,  \label{several}
\end{equation}
where $t_{0}$ is the starting time of the option contract. Note that after
an even number of mirrorings $S_{t}^{\ast }$ equals the value of the
underlying $S_{t}$ affected by a leveraging factor depending on the
historical values taken by the underlying at the mirroring times selected by
the holder. After an odd number of mirrorings we have also a path dependent
factor but in this case multiplying the inverse value of the underlying. In
Fig.(\ref{mirrorfig3}) we can see the effects of two consecutive mirrorings
over the sample path of Fig.(\ref{mirrorfig1}). 

Once it is clear what we mean by mirrorings we can define a mirror option as
following:

\begin{definition}
\label{def1}Given a European or American option with payoff $f$ we define
its mirror counterpart as the one with payoff $f^{\ast }$ given by 
\[
f^{\ast }\left( S\right) \equiv f\left( S^{\ast }\right) , 
\]
The holder of the mirror option has the right (not the obligation) at each
time between $t_{0}$ (start of the contract) and $T$ (maturity of the
contract) to perform mirrorings. The value $S_{t}^{\ast }$ of the mirror
underlying at time $t$ is given by Eq.(\ref{several}) where $t_{1}\leq
t_{2}\leq \cdots \leq t_{M}$ is the whole set of times where the holder has
chosen to perform mirrorings. Besides the right to perform mirrorings the
holder has all the rights associated to the standard option counterpart
(e.g. early exercise in the American case).
\end{definition}

Please note that we understand the holder is not allowed to mirror the past
history. Only future and therefore non-predictable history is affected by
the holder's mirrorings. Note also that the definition is sufficiently
general to accommodate mirror versions of plain-vanilla and exotic options
including the path-dependent case. However, we do not intend here to give a
static definition. We want instead to make clear which is the spirit of the
definition and therefore allow the reader to adapt it to other
possibilities. Just to give an example we can consider mirror exchanges or
rainbows \cite{exchange,rainbow} where we have a payoff that depends on
several underlyings and where the holder is allowed to perform independent
mirrorings on each of them.

From the definition it is now clear that depending on the holder ability to
foresee the tendency of the underlying the leveraging factor resulting from
judiciously chosen mirrorings can generate a much more convenient payoff
than the one obtained with the standard option counterpart. For example a
holder of a mirror call will try to perform mirrorings in order to maximize
the rise of $S^{\ast }$ and vice-versa for a mirror put. In the case of
options with more general payoffs the holder ability must be directed
towards the maximization of the payoff whatever it could be. In some sense
mirror options are reminiscent of chooser options \cite{chooser} with the
difference that here the holder `chooses' all the time during the contract.

\section{Valuating Mirror Options}

\label{valuating}For the valuation of mirror options we will assume the
simplest market model we can start with. Namely a one factor model
consisting of a single underlying $S$, paying the continuous dividend rate $%
\delta $, together with a riskless bonus $P.$ The pair of assets satisfy 
\begin{eqnarray}
\frac{dS_{t}}{S_{t}} &=&\mu dt+\sigma dW_{t},  \nonumber \\
\frac{dP_{t}}{P_{t}} &=&rdt,  \label{model}
\end{eqnarray}
where $dW_{t}$ is a Wiener process and where the drift $\mu $, volatility $%
\sigma $ and riskless interest rate $r$ are taken constant. We also assume
that there are no transactions costs. Therefore we will work with a complete
and efficient model without transaction costs. This model can be complicated
considerably and can be adapted to the particular underlying or sets of
underlyings of the corresponding option. Evidently correlations enter into
the game when several underlyings are involved. Moreover, the constancy of
the drift and volatility is no longer acceptable when the option is a
compounded one and therefore the underlying is another option \cite{compound}%
. Even for a single equity underlying one can consider multifactor models
with stochastic volatility and term structure \cite
{HullWhite,Scott2,Stein,Heston,Ghysels,Derman,HandN}, inefficient market
models \cite{Fama,Wang,Serva,Zhang,Bouchaud,Matacz,Masoliver},
jump-diffusion models \cite{Merton,Scott1,Aase,Kou}, models with transaction
costs \cite{Leland,Hodges,Dewynne} etc. However, in this article we want to
present a general valuation formulae obtained with the simplest hypothesis
leaving refinements and the treatment of particular cases to further
developments.

Using Eq.(\ref{several}) after $M$ mirrorings we can apply It\^{o} lemma 
\cite{Oksendal,Karatzas} to Eq.(\ref{model}) to obtain 
\[
\frac{dS_{t}^{\ast }}{S_{t}^{\ast }}=\mu ^{\ast }dt+\sigma ^{\ast }dW_{t}, 
\]
with 
\begin{eqnarray*}
\sigma ^{\ast } &=&\sigma , \\
\mu ^{\ast } &=&\left\{ 
\begin{array}{ccc}
\mu &  & M\quad even, \\ 
\sigma ^{2}-\mu &  & M\quad odd.
\end{array}
\right.
\end{eqnarray*}
From this fact one can be tempted to make the following incorrect argument:

The volatility of $S^{\ast }$ is unaffected by the mirroring operations made
by the holder and by definition \ref{def1} the payoff of the mirror option
is given by the same function that corresponds to the standard counterpart
but computed using $S^{\ast }$ instead of $S$. Moreover, since in the
valuation using a replicating self-financing portfolio the drift of the
underlying drops out from the calculations \cite{Cox} we can naively
conclude that the value of the mirror option is the same as the value of its
standard counterpart.

As the reader has surely noted the reason why the above argument is wrong is
that $S^{\ast }$ {\em is not the value of any tradable }\cite{Baxter,HandP}%
{\em . }Despite of that, we will see that the value of the mirror option is
closely related to the value of its standard counterpart, at least for a
family of payoffs where the valuation can be done in a closed form
(monotonic payoffs as we will see). Before starting the valuation we need
some preliminary results. For $z\in {\Bbb R}$ and $\Sigma \in {\Bbb R}^{+}$
we define 
\begin{eqnarray*}
\nu \left( z,\Sigma \right) &\equiv &\frac{1}{\sqrt{2\pi \Sigma }}%
\int_{-\infty }^{+\infty }f\left( e^{x}\right) e^{-\frac{1}{2\Sigma }\left(
x-z+\frac{1}{2}\Sigma \right) ^{2}}dx, \\
\nu \left( z,0\right) &\equiv &\lim_{\Sigma \rightarrow 0}\nu \left(
z,\Sigma \right) =f\left( e^{z}\right) ,
\end{eqnarray*}
where $f$ is the payoff function. Using the Chapman-Kolmogorov relation
satisfied by gaussians we immediately obtain 
\begin{equation}
\nu \left( z,\Sigma _{1}+\Sigma _{2}\right) =\frac{1}{\sqrt{2\pi \Sigma _{2}}%
}\int_{-\infty }^{+\infty }\nu \left( x,\Sigma _{1}\right) e^{-\frac{1}{%
2\Sigma _{2}}\left( x-z+\frac{1}{2}\Sigma _{2}\right) ^{2}}dx.  \label{tower}
\end{equation}
For $\phi ,\beta \in {\Bbb R}$ and $\Sigma _{1},\Sigma _{2}\in {\Bbb R}^{+}$
we also define 
\begin{equation}
\nu _{\phi ,\beta }\left( z,\Sigma _{1},\Sigma _{2}\right) \equiv \frac{1}{%
\sqrt{2\pi \Sigma _{2}}}\int_{-\infty }^{+\infty }\nu \left( \phi x+\beta
,\Sigma _{1}\right) e^{-\frac{1}{2\Sigma _{2}}\left( x-z+\frac{1}{2}\Sigma
_{2}\right) ^{2}}dx,  \label{deffb}
\end{equation}
then performing a change of variables and using Eq.(\ref{tower}) we obtain
the relation 
\begin{equation}
\nu _{\phi ,\beta }\left( z,\Sigma _{1},\Sigma _{2}\right) =\nu \left( \phi
z+\beta +\frac{\phi ^{2}-\phi }{2}\Sigma _{2},\Sigma _{1}+\phi ^{2}\Sigma
_{2}\right) .  \label{relat}
\end{equation}
Now we state the following simple result

\begin{lemma}
\label{lemma1}{\it The payoff }$f$ {{\it is a monotonic increasing
(decreasing) function if and only if} }$\forall \varepsilon \geq 0${\ {\it \ 
}}$\nu \left( \cdot ,\varepsilon \right) $ {\it is a monotonic increasing
(decreasing) function.}
\end{lemma}

{\bf Proof. }Suppose that $\forall \varepsilon \geq 0$ $\nu \left( \cdot
,\varepsilon \right) $ is a monotonic increasing (decreasing) function, then
in particular we have $\nu \left( \cdot ,0\right) =\left( f\circ \exp
\right) \left( \cdot \right) $ and since the exponential is a monotonic
increasing function we immediately obtain that the payoff is a monotonic
increasing (decreasing) function. Conversely, if the payoff{\it \ }$f$ is a
monotonic increasing (decreasing) function then $f\circ \exp $ is also a
monotonic increasing (decreasing) function and finally since $\nu \left(
\cdot ,\varepsilon \right) $ is the convolution of a positively defined
function (a gaussian in this case) with $f\circ \exp $ we immediately obtain
that $\nu \left( \cdot ,\varepsilon \right) $ is a monotonic increasing
(decreasing) function $\forall \varepsilon \geq 0$. $\blacksquare $

{\ }

{To perform the valuation we take time as discrete points and at the end we
will take the continuous limit. We define the time to maturity }$\tau $ as $%
\tau =n\varepsilon ${\ with} $n=0,1,2,\cdots ,\infty ${\ and} $\varepsilon ${%
\ arbitrary and constant. Since} $\varepsilon ${\ is arbitrary taking the
limits} $\varepsilon \rightarrow 0${, }$n\rightarrow \infty ${\ with} $\tau
=n\varepsilon ${\ fixed we obtain the value of the option for arbitrary time 
}$t=T-\tau ${.}

{Let us suppose that }at time $t=T-n\varepsilon $ the holder has performed a
certain number of mirrorings generating a virtual underlying value $%
S_{T-n\varepsilon }^{\ast }$ given by Eq.(\ref{several}). That value can be
very different from the real underlying value $S_{T-n\varepsilon }$ (note
also that the value $S_{T-n\varepsilon }^{\ast }$ is not affected by a
mirroring at time $T-n\varepsilon $). To begin with let us introduce some
notation. We define 
\begin{eqnarray*}
x_{t} &\equiv &\log \left( S_{t}\right) , \\
x_{t}^{\ast } &\equiv &\log \left( S_{t}^{\ast }\right) ,
\end{eqnarray*}
hence using Eq.(\ref{several}) we obtain the relation 
\begin{equation}
x_{t+\varepsilon }^{\ast }=\phi _{t}x_{t+\varepsilon }+\beta _{t},
\label{fibeta}
\end{equation}
where $\phi _{t}$ takes values $\pm 1$ and is equal to $\left( -1\right)
^{M\left( t\right) }$ with $M\left( t\right) $ the number of mirroring
decisions made by the holder up to time $t$ {\bf inclusive}. From Eq.(\ref
{several}) we can see that $\beta _{t}$ also depends on the history of
mirrorings up to time $t$ inclusive. Now we need to prove the following
result

\begin{lemma}
\label{lemma2} 
\[
x_{t}^{\ast }=\phi _{t}x_{t}+\beta _{t}, 
\]
\end{lemma}

{\bf Proof. }From Eq.(\ref{several}), when passing from time $t$ to time $%
t+\varepsilon $, we have the following four possibilities: 
\begin{equation}
\left. 
\begin{array}{llcl}
S_{t}^{\ast }=\gamma S_{t} & {\rm and~no~mirroring~at~time~}t & {\rm then} & 
S_{t+\varepsilon }^{\ast }=\gamma S_{t+\varepsilon } \\ 
S_{t}^{\ast }=\eta S_{t}^{-1} & {\rm and~mirroring~at~time~}t & {\rm then} & 
S_{t+\varepsilon }^{\ast }=\eta S_{t}^{-2}S_{t+\varepsilon }
\end{array}
\right\} \phi _{t}=1,  \label{fi1}
\end{equation}
and 
\begin{equation}
\left. 
\begin{array}{llcl}
S_{t}^{\ast }=\eta S_{t}^{-1} & {\rm and~no~mirroring~at~time~}t & {\rm then}
& S_{t+\varepsilon }^{\ast }=\eta S_{t+\varepsilon }^{-1} \\ 
S_{t}^{\ast }=\gamma S_{t} & {\rm and~mirroring~at~time~}t & {\rm then} & 
S_{t+\varepsilon }^{\ast }=\gamma S_{t}^{2}S_{t+\varepsilon }^{-1}
\end{array}
\right\} \phi _{t}=-1,  \label{fi-1}
\end{equation}
where $\gamma $ and $\eta $ are factors that depend on the values taken by
the underlying at mirroring times {\bf previous to }$t$ (see Eq.(\ref
{several})). Hence taking logarithms in Eqs.(\ref{fi1}) and (\ref{fi-1}) we
obtain 
\[
x_{t+\varepsilon }^{\ast }-x_{t}^{\ast }=\phi _{t}\left( x_{t+\varepsilon
}-x_{t}\right) ,
\]
and using Eq.(\ref{fibeta}) we immediately obtain 
\begin{eqnarray*}
x_{t}^{\ast } &=&x_{t+\varepsilon }^{\ast }+\phi _{t}\left(
x_{t}-x_{t+\varepsilon }\right)  \\
&=&\phi _{t}x_{t+\varepsilon }+\beta _{t}+\phi _{t}\left(
x_{t}-x_{t+\varepsilon }\right)  \\
&=&\phi _{t}x_{t}+\beta _{t}.
\end{eqnarray*}
$\blacksquare $

With this result and Eq.(\ref{relat}) we are ready to start. We will denote
the value of the mirror option at time to maturity $\tau $ by $V^{\ast
}\left( \tau \right) $. It is obvious that $V^{\ast }\left( \tau \right) $
also has a certain dependence on the history of the underlying up to time $%
T-\tau $. For example at maturity time we have 
\begin{equation}
V^{\ast }\left( 0\right) =f\left( S_{T}^{\ast }\right) =v\left( x_{T}^{\ast
},0\right) =v\left( \phi _{T-\varepsilon }x_{T}+\beta _{T-\varepsilon
},0\right) .  \label{atmat}
\end{equation}
where $\phi _{T-\varepsilon }$ and $\beta _{T-\varepsilon }$ are clearly
history dependent. In general to avoid raveling more the notation we will
assume such dependence implicitly. However, it is important to realize that
because of Eq.(\ref{atmat}) mirror options enter into the category of Asian
options with the peculiarity that the path dependence is dictated by the
holder with his/her mirroring decisions.

Hereafter we will analyze the particular case where the payoff function $f$
is monotonic. Denoting by $V_{+}^{\ast }$ ($V_{-}^{\ast }$) the value of the
mirror option whenever the payoff is monotonic increasing (decreasing) we
define 
\begin{equation}
V_{\pm }^{\ast }\left( \tau \right) \equiv e^{-r\tau }q_{\pm }^{\ast }\left(
\tau \right) ,  \label{redef}
\end{equation}
Now will demonstrate by induction the main result of this work, that is:

\begin{theorem}
\label{theorem1}Defining $\alpha \equiv r-\delta ,$ for the European case we
have 
\begin{equation}
q_{\pm }^{\ast }\left( \tau \right) =\nu \left( x_{T-\tau }^{\ast }+\left( 
\frac{\sigma ^{2}}{2}\pm \left| \alpha -\frac{\sigma ^{2}}{2}\right| \right)
\tau ,\sigma ^{2}\tau \right) .  \label{result}
\end{equation}
\end{theorem}

{\bf Proof. }Using Eqs.(\ref{atmat}) and (\ref{redef}) the result is
trivially true for $\tau =0$. Supposing it valid for $\tau =n\varepsilon $
we will demonstrate it for $\tau =\left( n+1\right) \varepsilon $ and
therefore for arbitrary $n$ by induction. According to the standard
procedure \cite{Hull,Baxter,HandP,Cox} for our market model (\ref{model}),
the value of a European claim at $\tau =\left( n+1\right) \varepsilon $ is
given by the $r-$discounted mean value of the claim at $\tau =n\varepsilon $
using the {\em risk neutral measure} ($\mu \rightarrow \alpha $). Let us for
a moment remember here the case of an American option. There the value of
the option was given by the maximum between that discounted value and the
value obtained by early exercise. Here, like in the American case, we have
to choose between two values, namely 
\begin{eqnarray}
q_{\pm }^{\ast }\left( \left( n+1\right) \varepsilon \right)
&=&\max_{mirror\left( \left( n+1\right) \varepsilon \right) }\left\{ \frac{1%
}{\sigma \sqrt{2\pi \varepsilon }}\int_{-\infty }^{+\infty
}dx_{T-n\varepsilon }q_{\pm }^{\ast }\left( n\varepsilon \right) \right. 
\nonumber \\
&&\left. \times \exp \left( -\frac{1}{2\sigma ^{2}\varepsilon }\left(
x_{T-n\varepsilon }-x_{T-\left( n+1\right) \varepsilon }-\alpha \varepsilon +%
\frac{1}{2}\sigma ^{2}\varepsilon \right) ^{2}\right) \right\} ,
\label{fund}
\end{eqnarray}
where $\max\limits_{mirror\left( \tau \right) }\left\{ \cdot \right\} $
means the maximum taken between the values of the expression between braces
evaluated with and without mirroring at time to maturity $\tau $. Note that
the risk neutral measure used in Eq.(\ref{fund}) is just the one arising
from no-arbitrage arguments based on the standard $\Delta $-hedging strategy
for our model (\ref{model}).

Using the inductive hypothesis given by Eq.(\ref{result}) along with Eq.(\ref
{fibeta}) we have 
\begin{eqnarray*}
q_{\pm }^{\ast }\left( \left( n+1\right) \varepsilon \right) 
&=&\max_{mirror\left( \left( n+1\right) \varepsilon \right) }\left\{
\int_{-\infty }^{+\infty }dx_{T-n\varepsilon }\exp \left( -\frac{1}{2\sigma
^{2}\varepsilon }\left( x_{T-n\varepsilon }-x_{T-\left( n+1\right)
\varepsilon }-\alpha \varepsilon +\frac{1}{2}\sigma ^{2}\varepsilon \right)
^{2}\right) \right.  \\
&&\left. \times \frac{1}{\sigma \sqrt{2\pi \varepsilon }}\nu \left( \phi
_{T-\left( n+1\right) \varepsilon }x_{T-n\varepsilon }+\beta _{T-\left(
n+1\right) \varepsilon }+\left( \frac{\sigma ^{2}}{2}\pm \left| \alpha -%
\frac{\sigma ^{2}}{2}\right| \right) n\varepsilon ,\sigma ^{2}n\varepsilon
\right) \right\} ,
\end{eqnarray*}
now using Eqs.(\ref{deffb}) and (\ref{relat}) we can write 
\begin{eqnarray*}
q_{\pm }^{\ast }\left( \left( n+1\right) \varepsilon \right) 
&=&\max_{mirror\left( \left( n+1\right) \varepsilon \right) }\left\{ \nu
\left( \phi _{T-\left( n+1\right) \varepsilon }x_{T-\left( n+1\right)
\varepsilon }+\beta _{T-\left( n+1\right) \varepsilon }\right. \right.  \\
&&+\left. \left. \left( \alpha \phi _{T-\left( n+1\right) \varepsilon
}+\left( \frac{\sigma ^{2}}{2}\pm \left| \alpha -\frac{\sigma ^{2}}{2}%
\right| \right) n+\frac{1-\phi _{T-\left( n+1\right) \varepsilon }}{2}\sigma
^{2}\right) \varepsilon ,\sigma ^{2}\left( n+1\right) \varepsilon \right)
\right\} ,
\end{eqnarray*}
and from Lemma \ref{lemma2} we have 
\begin{eqnarray*}
q_{\pm }^{\ast }\left( \left( n+1\right) \varepsilon \right) 
&=&\max_{mirror\left( \left( n+1\right) \varepsilon \right) }\left\{ \nu
\left( x_{T-\left( n+1\right) \varepsilon }^{\ast }+\left( \alpha \phi
_{T-\left( n+1\right) \varepsilon }+\left( \frac{\sigma ^{2}}{2}\pm \left|
\alpha -\frac{\sigma ^{2}}{2}\right| \right) n\right. \right. \right.  \\
&&\left. \left. \left. +\frac{1-\phi _{T-\left( n+1\right) \varepsilon }}{2}%
\sigma ^{2}\right) \varepsilon ,\sigma ^{2}\left( n+1\right) \varepsilon
\right) \right\} ,
\end{eqnarray*}
noting that $x_{t}^{\ast }$ is invariant under mirrorings at time $t$ we have

\begin{eqnarray*}
q_{\pm }^{\ast }\left( \left( n+1\right) \varepsilon \right) &=&\max \left\{
\nu \left( x_{T-\left( n+1\right) \varepsilon }^{\ast }+\frac{\sigma ^{2}}{2}%
\left( n+1\right) \varepsilon +\left( \alpha -\frac{\sigma ^{2}}{2}\right)
\varepsilon \pm \left| \alpha -\frac{\sigma ^{2}}{2}\right| n\varepsilon
,\sigma ^{2}\left( n+1\right) \varepsilon \right) ,\right. \\
&&\left. \nu \left( x_{T-\left( n+1\right) \varepsilon }^{\ast }+\frac{%
\sigma ^{2}}{2}\left( n+1\right) \varepsilon -\left( \alpha -\frac{\sigma
^{2}}{2}\right) \varepsilon \pm \left| \alpha -\frac{\sigma ^{2}}{2}\right|
n\varepsilon ,\sigma ^{2}\left( n+1\right) \varepsilon \right) \right\} .
\end{eqnarray*}
Finally, using the payoff monotonicity hypothesis and Lemma \ref{lemma1} we
obtain 
\[
q_{\pm }^{\ast }\left( \left( n+1\right) \varepsilon \right) =\nu \left(
x_{T-\left( n+1\right) \varepsilon }^{\ast }+\left( \frac{\sigma ^{2}}{2}\pm
\left| \alpha -\frac{\sigma ^{2}}{2}\right| \right) \left( n+1\right)
\varepsilon ,\sigma ^{2}\left( n+1\right) \varepsilon \right) . 
\]
{\ }$\blacksquare $

Now we can trivially take the continuous limit $\varepsilon \rightarrow 0${, 
}$n\rightarrow \infty ${\ with} $\tau =n\varepsilon ${\ fixed. In this limit
and using Eq.(\ref{redef}) we obtain the value of the mirror option for
general monotonic payoff functions as} 
\begin{equation}
V_{\pm }^{\ast }\left( \tau \right) =e^{-r\tau }\nu \left( x_{T-\tau }^{\ast
}+\left( \frac{\sigma ^{2}}{2}\pm \left| r-\delta -\frac{\sigma ^{2}}{2}%
\right| \right) \tau ,\sigma ^{2}\tau \right) ,  \label{finalresult}
\end{equation}
This is a remarkable result, remembering that the value $V\left( \tau
\right) $ for the standard European option is given by 
\begin{equation}
V\left( \tau \right) =e^{-r\tau }\nu \left( x_{T-\tau }+\left( r-\delta
\right) \tau ,\sigma ^{2}\tau \right) ,  \label{standard}
\end{equation}
we immediately note that at the beginning of the contract ($x_{t_{0}}^{\ast
}=x_{t_{0}}$) if we have a monotonic increasing payoff (like in a call for
example) we have that the mirror option has the same (for $r-\delta \geq
\sigma ^{2}/2$) or bigger (for $r-\delta \leq \sigma ^{2}/2$) value than its
standard counterpart. Analogously, if we have a monotonic decreasing payoff
(like in a put for example) we have that the mirror option has the same (for 
$r-\delta \leq \sigma ^{2}/2$) or bigger (for $r-\delta \geq \sigma ^{2}/2$)
value than its standard counterpart. Thus we have arrived to the conclusion
that, at the beginning of the contract, the deviation of the value of the
mirror option from that of its standard counterpart depends critically on
the interplay between dividends, interest rates and volatility. Note also
that here dividends appear in a non trivial way, that is the rule to
incorporate continuous dividends given by $S_{t}\rightarrow S_{t}\exp \left(
-\delta \left( T-t\right) \right) $ is not valid for mirror options. Another
point that should be clear is that even though we have derived our
expressions for the monotonic payoff case, we can easily carry on the
demonstration of Theorem \ref{theorem1} in the limit case where $\sigma =%
\sqrt{2\left( r-\delta \right) }$ {\em for arbitrary payoff functions}. In
this `zero-measure' case we obtain that Eq.(\ref{finalresult}) remains valid
and therefore at the beginning of the contract we have 
\[
V^{\ast }\left( T-t_{0}\right) =V\left( T-t_{0}\right) \qquad \left( \sigma
=\sigma _{c}\right) ,
\]
where we define the critical volatility $\sigma _{c}$ as 
\[
\sigma _{c}=\sqrt{2\left( r-\delta \right) }.
\]
With this definition we can write Eq.(\ref{finalresult}) in the more
suggestive way 
\[
V_{\pm }^{\ast }\left( \tau \right) =e^{-r\tau }\nu \left( x_{T-\tau }^{\ast
}+\left( r-\delta \right) \tau +\left( \sigma ^{2}-\sigma _{c}^{2}\right)
_{\pm }\tau ,\sigma ^{2}\tau \right) ,
\]
where 
\[
\left( x\right) _{\pm }\equiv \left\{ 
\begin{array}{cl}
x & {\rm if~sign}\left( x\right) =\pm 1, \\ 
0 & {\rm otherwise}.
\end{array}
\right. 
\]
Another point to remark is that now we can clearly see the Asian
characteristics of mirror options. From the payoff definition of mirror
options we already knew that they were Asian options, but now it is clear
also that we have a dependence on $S_{t}^{\ast }$ for {\em any} time $t$
during the contract. Moreover, from Eq.(\ref{several}), we can interpret $%
S_{t}^{\ast }$ as a kind of weighted geometric average in a set of
`mirrored' values up to time $t.$ This set is, by definition, dictated by
the holder with his/her mirrorings decisions at each time up to $t$.

To illustrate these results we will consider some examples, but before that
let us clarify some potentially confusing language. It is standard in
financial language to talk about a {\em long} option position with payoff $f$
and about a {\em short} option position with payoff $-f$. Here we will
continue with this nomenclature but bearing in mind that unlike what happens
with standard contracts a {\em short} mirror option{\em \ }position {\em is
not} equivalent to a {\em sold} mirror option position. In the last case we
are taking about the position of the {\em issuer} of the long mirror option
that is {\em not} equivalent to any position of a {\em holder} because it is
only the later who takes the mirroring decisions. Because of this fact
holding both a {\em short} and a ${\em long}$ mirror option position has in
general some non-vanishing value. For the monotonic payoff case this value
is 
\begin{eqnarray}
\left( V_{long}^{\ast }+V_{short}^{\ast }\right) \left( T-t\right)
&=&e^{-r\left( T-t\right) }\left\{ \nu \left( x_{t}^{\left( long\right) \ast
}+\left( \frac{\sigma ^{2}}{2}\pm \left| r-\delta -\frac{\sigma ^{2}}{2}%
\right| \right) \tau ,\sigma ^{2}\tau \right) \right.  \nonumber \\
&&\left. -\nu \left( x_{t}^{\left( short\right) \ast }+\left( \frac{\sigma
^{2}}{2}\mp \left| r-\delta -\frac{\sigma ^{2}}{2}\right| \right) \tau
,\sigma ^{2}\tau \right) \right\} ,  \label{longshort}
\end{eqnarray}

where the upper (lower) sign corresponds to the case where the payoff $f$
for the long position is monotonic increasing (decreasing). Note that in
this case the holder has the right to perform a certain set of mirrorings
for his/her long position and a {\em different} set of mirrorings for
his/her short position and therefore as time goes by he/she will have in
general a mirror path $x_{t}^{\left( long\right) \ast }$ for the long
position and a another mirror path $x_{t}^{\left( short\right) \ast }$ for
the short position. At the beginning of the contract ($t=t_{0}$) we have $%
x_{t_{0}}^{\left( long\right) \ast }=x_{t_{0}}^{\left( short\right) \ast
}=x_{t_{0}}$ and therefore using Lemma \ref{lemma1} we can clearly see that
the value given by Eq.(\ref{longshort}) at this time is always {\em positive
definite}. Moreover the unique possibility for this value to be zero is the
limit case where $\sigma =\sigma _{c}$ which is the case where all mirror
options take the same value as their standard counterparts.

\subsection{The European Mirror Call}

A European long (short) call has the payoff $f\left( S\right) =\pm \left(
S-K\right) _{+}$where the upper (lower) sign corresponds to the long (short)
position. Modeled with Eq.(\ref{model}) the long (short) call has the value $%
V^{\left( \pm \right) }\left( \tau \right) $ given by the Black-Scholes
formula \cite{Hull} 
\[
V^{\left( \pm \right) }\left( \tau \right) =\pm e^{-r\tau }\left\{ F_{\tau
}\Phi \left( \frac{\ln \left( \frac{F_{\tau }}{K}\right) +\frac{1}{2}\sigma
^{2}\tau }{\sigma \sqrt{\tau }}\right) -K\Phi \left( \frac{\ln \left( \frac{%
F_{\tau }}{K}\right) -\frac{1}{2}\sigma ^{2}\tau }{\sigma \sqrt{\tau }}%
\right) \right\} , 
\]
with 
\[
F_{\tau }\equiv S_{T-\tau }e^{\left( r-\delta \right) \tau }. 
\]
Hence from Eq.(\ref{finalresult}) and since in this case the payoff is a
monotonic increasing (decreasing) function of $S$ we immediately obtain the
value of the corresponding European mirror call as 
\[
V^{\left( \pm \right) \ast }\left( \tau \right) =\pm e^{-r\tau }\left\{
F_{\tau }^{\left( \pm \right) \ast }\Phi \left( \frac{\ln \left( \frac{%
F_{\tau }^{\left( \pm \right) \ast }}{K}\right) +\frac{1}{2}\sigma ^{2}\tau 
}{\sigma \sqrt{\tau }}\right) -K\Phi \left( \frac{\ln \left( \frac{F_{\tau
}^{\left( \pm \right) \ast }}{K}\right) -\frac{1}{2}\sigma ^{2}\tau }{\sigma 
\sqrt{\tau }}\right) \right\} , 
\]
with 
\[
F_{\tau }^{\left( \pm \right) \ast }\equiv S_{T-\tau }^{\ast }e^{\left( 
\frac{\sigma ^{2}}{2}\pm \left| r-\delta -\frac{\sigma ^{2}}{2}\right|
\right) \tau }. 
\]
Note that necessarily one of the two mirror positions (short or long) has
the same value as its standard counterpart at the beginning of the contract.

\subsection{The European Mirror Put}

A European long (short) put has the payoff $f\left( S\right) =\pm \left(
K-S\right) _{+}$ where the upper (lower) sign corresponds to the long
(short) position. Modeled with Eq.(\ref{model}) the long (short) put has the
value $V^{\left( \pm \right) }\left( \tau \right) $ given by the
Black-Scholes formula \cite{Hull} 
\[
V^{\left( \pm \right) }\left( \tau \right) =\pm e^{-r\tau }\left\{ K\Phi
\left( -\frac{\ln \left( \frac{F_{\tau }}{K}\right) -\frac{1}{2}\sigma
^{2}\tau }{\sigma \sqrt{\tau }}\right) -F_{\tau }\Phi \left( -\frac{\ln
\left( \frac{F_{\tau }}{K}\right) +\frac{1}{2}\sigma ^{2}\tau }{\sigma \sqrt{%
\tau }}\right) \right\} , 
\]
with 
\[
F_{\tau }\equiv S_{T-\tau }e^{\left( r-\delta \right) \tau }. 
\]
Hence from Eq.(\ref{finalresult}) and since in this case the payoff is a
monotonic decreasing (increasing) function of $S$ we immediately obtain the
value of the corresponding European mirror put as 
\[
V^{\left( \pm \right) \ast }\left( \tau \right) =\pm e^{-r\tau }\left\{
K\Phi \left( -\frac{\ln \left( \frac{F_{\tau }^{\left( \pm \right) \ast }}{K}%
\right) -\frac{1}{2}\sigma ^{2}\tau }{\sigma \sqrt{\tau }}\right) -F_{\tau
}^{\ast }\Phi \left( -\frac{\ln \left( \frac{F_{\tau }^{\left( \pm \right)
\ast }}{K}\right) +\frac{1}{2}\sigma ^{2}\tau }{\sigma \sqrt{\tau }}\right)
\right\} , 
\]
with 
\[
F_{\tau }^{\left( \pm \right) \ast }\equiv S_{T-\tau }^{\ast }e^{\left( 
\frac{\sigma ^{2}}{2}\mp \left| r-\delta -\frac{\sigma ^{2}}{2}\right|
\right) \tau }. 
\]
Note again that necessarily one of the two mirror positions (short or long)
has the same value as its standard counterpart at the beginning of the
contract

\subsection{The Mirror Forward}

A long (short) forward contract has the payoff $f\left( S\right) =\pm \left(
S-K\right) $ where the upper (lower) sign corresponds to the long (short)
position. Using simple no-arbitrage arguments \cite{Hull} we obtain its
value $V^{\left( \pm \right) }\left( \tau \right) $ given by 
\[
V^{\left( \pm \right) }\left( \tau \right) =\pm e^{-r\tau }\left( F_{\tau
}-K\right) , 
\]
with 
\[
F_{\tau }\equiv S_{T-\tau }e^{\left( r-\delta \right) \tau }. 
\]
Hence from Eq.(\ref{finalresult}) and since in this case the payoff is a
monotonic increasing (decreasing) function of $S$ we immediately obtain the
value of the corresponding mirror forward contract as 
\[
V^{\left( \pm \right) \ast }\left( \tau \right) =\pm e^{-r\tau }\left(
F_{\tau }^{\left( \pm \right) \ast }-K\right) , 
\]
with 
\begin{equation}
F_{\tau }^{\left( \pm \right) \ast }\equiv S_{T-\tau }^{\ast }e^{\left( 
\frac{\sigma ^{2}}{2}\pm \left| r-\delta -\frac{\sigma ^{2}}{2}\right|
\right) \tau }.  \label{forward}
\end{equation}
Like what happens with the standard forward contract here we can also choose
the strike $K$ (the forward price) in order to make the price of the forward
contract zero. Note also that from Eq.(\ref{forward}) the forward prices for
the long and short positions are different. However, as we have seen, one of
them is necessarily equal to the standard value.

We can use this simple result to exemplify the positive definite value
arising from holding both a long and a short position at the beginning of
the contract. According to Eq.(\ref{longshort}) we have 
\[
\left( V^{\left( +\right) \ast }+V^{\left( -\right) \ast }\right) \left(
\tau \right) =e^{-r\tau }\left( S_{T-\tau }^{\left( long\right) \ast
}e^{\left( \frac{\sigma ^{2}}{2}+\left| r-\delta -\frac{\sigma ^{2}}{2}%
\right| \right) \tau }-S_{T-\tau }^{\left( short\right) \ast }e^{\left( 
\frac{\sigma ^{2}}{2}-\left| r-\delta -\frac{\sigma ^{2}}{2}\right| \right)
\tau }\right) , 
\]
where $S_{T-\tau }^{\left( long\right) \ast }$ ($S_{T-\tau }^{\left(
short\right) \ast }$) is the value of the mirror underlying for the long
(short) position. At time $t_{0}$ (the beginning of the contract) we have $%
S_{t_{0}}^{\left( long\right) \ast }=S_{t_{0}}^{\left( short\right) \ast
}=S_{t_{0}}$ so for this time we have 
\begin{equation}
\left( V^{\left( +\right) \ast }+V^{\left( -\right) \ast }\right) \left(
T-t_{0}\right) =2e^{\left( \frac{\sigma ^{2}}{2}-r\right) \left(
T-t_{0}\right) }\sinh \left( \left| r-\delta -\frac{\sigma ^{2}}{2}\right|
\left( T-t_{0}\right) \right) S_{t_{0}},  \label{lsfuture}
\end{equation}
which is clearly positive definite as we have already shown in general. From
the holder's perspective the payoff at maturity will be 
\[
S_{T}^{\left( long\right) \ast }-S_{T}^{\left( short\right) \ast }, 
\]
with the break-even given by 
\begin{equation}
\frac{S_{T}^{\left( long\right) \ast }-S_{T}^{\left( short\right) \ast }}{%
S_{t_{0}}}=2e^{\frac{\sigma ^{2}}{2}\left( T-t_{0}\right) }\sinh \left(
\left| r-\delta -\frac{\sigma ^{2}}{2}\right| \left( T-t_{0}\right) \right) .
\label{breakevent}
\end{equation}

\section{Hedging Mirror Options}

\label{hedging}In this section our aim is to show a very important feature
appearing in the hedging of mirror options that is highly relevant when
transaction costs and illiquidity of the market are taken into account. This
in turn will provide us with another way to envisage mirror options as a
kind of `swaps' of transaction costs between the holder and the issuer.

To calculate the value of mirror options in section \ref{valuating} we have
used the neutral risk probability arising in a $\Delta $-hedging strategy.
Let us now calculate explicitly such $\Delta $ from result (\ref{finalresult}%
). We define 
\begin{eqnarray*}
\Delta  &\equiv &\frac{\partial V}{\partial S}, \\
\Delta _{\pm }^{\ast } &\equiv &\frac{\partial V_{\pm }^{\ast }}{\partial S},
\end{eqnarray*}
hence using Eqs.(\ref{finalresult}) and (\ref{standard}) we obtain 
\begin{eqnarray}
\Delta  &=&e^{-r\tau }\frac{1}{S}\frac{\partial \nu }{\partial z}\left(
w,\sigma ^{2}\tau \right) ,  \nonumber \\
\Delta _{\pm }^{\ast } &=&e^{-r\tau }\frac{1}{S}\frac{\partial \nu }{%
\partial z}\left( w_{\pm }^{\ast },\sigma ^{2}\tau \right) \times \frac{S}{%
S^{\ast }}\frac{\partial S^{\ast }}{\partial S},  \label{delta}
\end{eqnarray}
with 
\begin{eqnarray*}
w &=&x_{T-\tau }+\left( r-\delta \right) \tau , \\
w_{\pm }^{\ast } &=&x_{T-\tau }^{\ast }+\left( \frac{\sigma ^{2}}{2}\pm
\left| r-\delta -\frac{\sigma ^{2}}{2}\right| \right) \tau ,
\end{eqnarray*}
and 
\begin{equation}
\frac{S}{S^{\ast }}\frac{\partial S^{\ast }}{\partial S}=\left( -1\right)
^{M\left( t\right) }.  \label{mirrorsign}
\end{equation}
The most important point to note in Eq.(\ref{delta}) is {\em not} the
difference between $w$ and $w_{\pm }^{\ast }$, that was already discussed in
the previous sections, but the sign flip given by Eq.(\ref{mirrorsign}). In
other words we have obtained that, from the hedger perspective, a mirroring
made by the holder implies a change in the hedging position in the
underlying from a given $\Delta _{\pm }^{\ast }$ to the opposite one. This
result is in fact very intuitive; since mirror options were created as
instruments allowing the holder costless changes in his/her position it is
natural to have the cost of these changes transferred to the issuer (zero in
our model). This is, again, very reasonable because issuers are in general
large investment banks or financial institutions that are exposed to much
lower transaction costs than individual speculators and therefore can absorb
such changes in position much more easily. Moreover, such financial
institutions do not hedge single instruments individually because the impact
of transactions costs is highly non-linear \cite{Dewynne}. As it is well
known, it is the {\em complete portfolio} of options and other products what
is hedged as a whole. Hedging each product individually is not realistic
because such policy do not profit from cancellations between different
positions and reductions of transaction costs due to economies of scale.
Therefore the real cost of hedging mirror options is minimized with respect
to the total premium received when large quantities of mirror options are
issued in the marker hedging them as a whole along with the rest of products
in the portfolio. However, in situations of large illiquidity (like those
appearing in crashes or rallies) collective behavior of mirror option
holders can lead to large changes in delta that can not be easily canceled
out. Because of that is not unrealistic to conceive that in a real market of
mirror options a risk premium depending in some way on the number of allowed
mirrorings can appear. In order to admit such correction we need to know the
value of mirror options when only a finite number of mirrorings is allowed.
This is the subject of the next section.

\section{Mirror options with a finite number of allowed mirrorings}

\label{finiteM}In this section we will obtain the value of a mirror option
for the case where the holder is allowed to perform $M$ mirrorings at most.
Again we will be able to obtain closed expressions for the monotonic payoff
case or, at most, for arbitrary payoffs in the critical point $\sigma
=\sigma _{c}$. With the same notational conventions we used in section \ref
{valuating} we define the value of the mirror option with at most $M$
allowed mirrorings as 
\[
V_{\pm }^{\ast }\left( \tau ,M\right) \equiv e^{-r\tau }q_{\pm }^{\ast
}\left( \tau ,M\right) . 
\]
In this case we have

\begin{theorem}
\label{theorem2} Defining $\alpha \equiv r-\delta ,$ for the European case,
where the holder is allowed to perform $M$ mirrorings at most, we have 
\begin{equation}
q_{\pm }^{\ast }\left( \tau ,M\right) =\left\{ 
\begin{array}{lcl}
\nu \left( x_{T-\tau }^{\ast }+\left( \frac{\sigma ^{2}}{2}\pm \left| \alpha
-\frac{\sigma ^{2}}{2}\right| \right) \tau ,\sigma ^{2}\tau \right) &  & 
{\rm if}~M\left( T-\tau -\varepsilon \right) <M, \\ 
\nu \left( x_{T-\tau }^{\ast }+\left( \frac{\sigma ^{2}}{2}+\left( -1\right)
^{M}\left( \alpha -\frac{\sigma ^{2}}{2}\right) \right) \tau ,\sigma
^{2}\tau \right) &  & {\rm if}~M\left( T-\tau -\varepsilon \right) =M,
\end{array}
\right.  \label{result2}
\end{equation}
where with $M\left( T-\tau -\varepsilon \right) <M$ we mean that at time $%
T-\tau $ the holder has at least one mirroring to perform and with $M\left(
T-\tau -\varepsilon \right) =M$ we mean that at time $T-\tau $ the holder
has exhausted all his/her allowed mirrorings.
\end{theorem}

{\bf Proof. }Using Eqs.(\ref{atmat}) and (\ref{redef}) the result is
trivially true for $\tau =0$. Again, supposing it valid for $\tau
=n\varepsilon $ we will demonstrate it for $\tau =\left( n+1\right)
\varepsilon $ and therefore for arbitrary $n$ by induction. Using the same
arguments of theorem \ref{theorem1} we obtain 
\[
q_{\pm }^{\ast }\left( \left( n+1\right) \varepsilon ,M\right) =\left\{ 
\begin{array}{ccl}
a_{\pm }^{\ast }\left( \left( n+1\right) \varepsilon ,M\right)  &  & {\rm if}%
~M\left( T-\left( n+2\right) \varepsilon \right) <M{\rm ,} \\ 
b_{\pm }^{\ast }\left( \left( n+1\right) \varepsilon ,M\right)  &  & {\rm if}%
~M\left( T-\left( n+2\right) \varepsilon \right) =M{\rm ,}
\end{array}
\right. 
\]
where 
\begin{eqnarray*}
a_{\pm }^{\ast }\left( \left( n+1\right) \varepsilon ,M\right) 
&=&\max_{mirror\left( \left( n+1\right) \varepsilon \right) }\left\{ \frac{1%
}{\sigma \sqrt{2\pi \varepsilon }}\int_{-\infty }^{+\infty
}dx_{T-n\varepsilon }q_{\pm }^{\ast }\left( n\varepsilon ,M\right) \right. 
\\
&&\left. \times \exp \left( -\frac{1}{2\sigma ^{2}\varepsilon }\left(
x_{T-n\varepsilon }-x_{T-\left( n+1\right) \varepsilon }-\alpha \varepsilon +%
\frac{1}{2}\sigma ^{2}\varepsilon \right) ^{2}\right) \right\} ,
\end{eqnarray*}
and 
\begin{eqnarray*}
b_{\pm }^{\ast }\left( \left( n+1\right) \varepsilon ,M\right)  &=&\frac{1}{%
\sigma \sqrt{2\pi \varepsilon }}\int_{-\infty }^{+\infty }dx_{T-n\varepsilon
}b_{\pm }^{\ast }\left( n\varepsilon ,M\right)  \\
&&\times \exp \left( -\frac{1}{2\sigma ^{2}\varepsilon }\left(
x_{T-n\varepsilon }-x_{T-\left( n+1\right) \varepsilon }-\alpha \varepsilon +%
\frac{1}{2}\sigma ^{2}\varepsilon \right) ^{2}\right) .
\end{eqnarray*}
Using the inductive hypothesis given by Eq.(\ref{result2}) along with Eqs.(%
\ref{deffb}), (\ref{relat}), (\ref{fibeta}) and Lemma \ref{lemma2} we obtain 
\begin{eqnarray*}
b_{\pm }^{\ast }\left( \left( n+1\right) \varepsilon ,M\right)  &=&\frac{1}{%
\sigma \sqrt{2\pi \varepsilon }}\int_{-\infty }^{+\infty }dx_{T-n\varepsilon
}\exp \left( -\frac{1}{2\sigma ^{2}\varepsilon }\left( x_{T-n\varepsilon
}-x_{T-\left( n+1\right) \varepsilon }-\alpha \varepsilon +\frac{1}{2}\sigma
^{2}\varepsilon \right) ^{2}\right)  \\
&&\times \nu \left( \phi _{T-\left( n+1\right) \varepsilon
}x_{T-n\varepsilon }+\beta _{T-\left( n+1\right) \varepsilon }+\left( \frac{%
\sigma ^{2}}{2}+\left( -1\right) ^{M}\left( \alpha -\frac{\sigma ^{2}}{2}%
\right) \right) n\varepsilon ,\sigma ^{2}n\varepsilon \right)  \\
&=&\nu \left( x_{T-\left( n+1\right) \varepsilon }^{\ast }+\left( \frac{%
\sigma ^{2}}{2}+\left( -1\right) ^{M}\left( \alpha -\frac{\sigma ^{2}}{2}%
\right) \right) \left( n+1\right) \varepsilon ,\sigma ^{2}\left( n+1\right)
\varepsilon \right) .
\end{eqnarray*}
where it is clear that when considering $b_{\pm }^{\ast }\left( \tau
,M\right) $ we have $\phi _{T-\tau }=\left( -1\right) ^{M}.$ Note that so
far we have not used the payoff monotonicity hypothesis. For $a_{\pm }^{\ast
}$ using the inductive hypothesis given by Eq.(\ref{result2}) along with Eq.(%
\ref{fibeta}) we have 
\begin{eqnarray*}
a_{\pm }^{\ast }\left( \left( n+1\right) \varepsilon ,M\right) 
&=&\max_{mirror\left( \left( n+1\right) \varepsilon \right) }\left\{
\int_{-\infty }^{+\infty }\nu \left( \phi _{T-\left( n+1\right) \varepsilon
}x_{T-n\varepsilon }+\beta _{T-\left( n+1\right) \varepsilon }+\left( \frac{%
\sigma ^{2}}{2}+\gamma _{\pm }\right) n\varepsilon ,\sigma ^{2}n\varepsilon
\right) \right.  \\
&&\left. \times \frac{1}{\sigma \sqrt{2\pi \varepsilon }}\exp \left( -\frac{1%
}{2\sigma ^{2}\varepsilon }\left( x_{T-n\varepsilon }-x_{T-\left( n+1\right)
\varepsilon }-\alpha \varepsilon +\frac{1}{2}\sigma ^{2}\varepsilon \right)
^{2}\right) dx_{T-n\varepsilon }\right\} ,
\end{eqnarray*}
where 
\[
\gamma _{\pm }=\left\{ 
\begin{array}{ccl}
\pm \left| \alpha -\frac{\sigma ^{2}}{2}\right|  &  & {\rm if}~M\left(
T-\left( n+1\right) \varepsilon \right) <M, \\ 
\left( -1\right) ^{M}\left( \alpha -\frac{\sigma ^{2}}{2}\right)  &  & {\rm %
if}~M\left( T-\left( n+1\right) \varepsilon \right) =M,
\end{array}
\right. 
\]
hence using Eqs.(\ref{deffb}), (\ref{relat}) and Lemma \ref{lemma2} we
obtain 
\begin{eqnarray*}
a_{\pm }^{\ast }\left( \left( n+1\right) \varepsilon ,M\right) 
&=&\max_{mirror\left( \left( n+1\right) \varepsilon \right) }\left\{ \nu
\left( x_{T-\left( n+1\right) \varepsilon }^{\ast }+\frac{\sigma ^{2}}{2}%
\left( n+1\right) \varepsilon \right. \right.  \\
&&\left. \left. +\left( \phi _{T-\left( n+1\right) \varepsilon }\left(
\alpha -\frac{\sigma ^{2}}{2}\right) +\gamma _{\pm }n\right) \varepsilon
,\sigma ^{2}\left( n+1\right) \varepsilon \right) \right\} .
\end{eqnarray*}
Now we have  to analyze two cases. If $M\left( T-\left( n+2\right)
\varepsilon \right) =M-1$, remembering that $\phi _{t}=\left( -1\right)
^{M\left( t\right) }$ and using Lemma \ref{lemma1} we obtain 
\begin{eqnarray*}
a_{\pm }^{\ast }\left( \left( n+1\right) \varepsilon ,M\right)  &=&\max
\left\{ \nu \left( x_{T-\left( n+1\right) \varepsilon }^{\ast }+\frac{\sigma
^{2}}{2}\left( n+1\right) \varepsilon +\left( -1\right) ^{M}\left( \alpha -%
\frac{\sigma ^{2}}{2}\right) \left( n+1\right) \varepsilon ,\sigma
^{2}\left( n+1\right) \varepsilon \right) ,\right.  \\
&&\left. \nu \left( x_{T-\left( n+1\right) \varepsilon }^{\ast }+\frac{%
\sigma ^{2}}{2}\left( n+1\right) \varepsilon -\left( -1\right) ^{M}\left(
\alpha -\frac{\sigma ^{2}}{2}\right) \varepsilon \pm \left| \alpha -\frac{%
\sigma ^{2}}{2}\right| n\varepsilon ,\sigma ^{2}\left( n+1\right)
\varepsilon \right) \right\}  \\
&=&\nu \left( x_{T-\left( n+1\right) \varepsilon }^{\ast }+\left( \frac{%
\sigma ^{2}}{2}\pm \left| \alpha -\frac{\sigma ^{2}}{2}\right| \right)
\left( n+1\right) \varepsilon ,\sigma ^{2}\left( n+1\right) \varepsilon
\right) ,
\end{eqnarray*}
and finally if $M\left( T-\left( n+2\right) \varepsilon \right) <M-1$ we
obtain 
\begin{eqnarray*}
a_{\pm }^{\ast }\left( \left( n+1\right) \varepsilon ,M\right)  &=&\max
\left\{ \nu \left( x_{T-\left( n+1\right) \varepsilon }^{\ast }+\frac{\sigma
^{2}}{2}\left( n+1\right) \varepsilon +\left( \alpha -\frac{\sigma ^{2}}{2}%
\right) \varepsilon \pm \left| \alpha -\frac{\sigma ^{2}}{2}\right|
n\varepsilon ,\sigma ^{2}\left( n+1\right) \varepsilon \right) ,\right.  \\
&&\left. \nu \left( x_{T-\left( n+1\right) \varepsilon }^{\ast }+\frac{%
\sigma ^{2}}{2}\left( n+1\right) \varepsilon -\left( \alpha -\frac{\sigma
^{2}}{2}\right) \varepsilon \pm \left| \alpha -\frac{\sigma ^{2}}{2}\right|
n\varepsilon ,\sigma ^{2}\left( n+1\right) \varepsilon \right) \right\}  \\
&=&\nu \left( x_{T-\left( n+1\right) \varepsilon }^{\ast }+\left( \frac{%
\sigma ^{2}}{2}\pm \left| \alpha -\frac{\sigma ^{2}}{2}\right| \right)
\left( n+1\right) \varepsilon ,\sigma ^{2}\left( n+1\right) \varepsilon
\right) .
\end{eqnarray*}

$\blacksquare $

Note that as expected we have 
\[
q_{\pm }^{\ast }\left( \tau ,\infty \right) =q_{\pm }^{\ast }\left( \tau
\right) . 
\]
\qquad Moreover, from Eq.(\ref{result2}) we can see (at least for the
monotonic payoff case) that the value of the mirror option {\em before the
holder has exhausted his/her last mirroring} do not depend on the allowed
number of mirrorings $M.$ Immediately after the last allowed mirroring is
executed the value of the mirror option always falls to a lower or equal
value. The discontinuity arises whenever 
\[
\pm \left| \alpha -\frac{\sigma ^{2}}{2}\right| \neq \left( -1\right)
^{M}\left( \alpha -\frac{\sigma ^{2}}{2}\right) , 
\]
where the upper (lower) sign corresponds to the monotonic increasing
(decreasing) payoff case. For example we will have a discontinuous fall
immediately after the last mirroring in a long mirror call if $M$ is even
(odd) and $\sigma >\sigma _{c}$ ($\sigma <\sigma _{c}$). Note also that the
expression given by Eq.(\ref{result2}) in the trivial case $M=0$ reduces to
the standard result (\ref{standard}).

So far the reader may find that our results are counterintuitive since at
the beginning of the contract two mirror options differing only in the
number of allowed mirrorings have exactly the same value (if at least 1
mirroring is allowed in both). However in our simplified model our results
are correct and must be taken as a base for further developments
incorporating the corresponding corrections (see the conclusions for a brief
discussion).

\section{Conclusions and Outlook}

\label{conclusions}In this work we have created a new family of options
specially crafted to satisfy the necessities of those speculators who are
convinced of their abilities to call the market and do not want to assume
the transaction costs of daily trading. Moreover, we were able to obtain
general formulae for the value of these options in the case where the payoff
is a monotonic function (which is the case in calls, puts, futures, spreads,
asset or nothing options \cite{binary}, etc.). Moreover, for the general
payoff case we have identified a critical volatility $\sigma _{c}=\sqrt{%
2\left( r-\delta \right) }$ for which the value of {\em all mirror options,}
at the beginning of the contract, equals that of their standard counterparts.

We conclude our work with two sets of general remarks. The first set regards
general aspects of the pricing of mirror options within the hypothesis of
our model. The second set considers the impact other models may have on our
results.

\bigskip

From simple financial arguments we know that holding a long call and a long
put with the same strike is equivalent to holding a long straddle. Note
however that holding a long mirror call and a long mirror put with the same
strike is not equivalent to holding a long mirror straddle. This is because
in the last case the holder has the right to perform a set of mirrorings for
the call and another different set of mirrorings for the put. This is
evidently not equivalent to the situation of holding a single mirror
straddle. This fact is rather general in the sense that care must be taken
when considering the valuation of synthetic options made of mirror options.

Since one long mirror call plus one long mirror put is not one long mirror
straddle, which is the price of a long mirror straddle then? Or more
generally: which is the price of a mirror option with non-monotonic payoff?
The answer can be obtained following the steps of valuation of sections \ref
{valuating} and \ref{finiteM} with the difference that since the straddle
has not monotonic payoff we cannot use this property and because of that
only a numerical method (the binomial tree for example) can be applied in
principle. In general, the difficulty of valuating mirror options with not
monotonic payoffs is equivalent to the difficulty of valuating American
options where free boundary problems arise \cite{Wilmott}. Here the free
boundary is given, for each time $t,$ by the value $S_{t}^{\ast }$ for which
mirroring is equivalent to no mirroring; in a complete analogous way to the
free boundary arising in the American option case given by the equivalence
of early exercise or not.

In this work we have valuated European mirror options, but what about
American mirror options? Again, the valuation can be obtained using the
steps of sections \ref{valuating} and \ref{finiteM} but now at each time
stage we have to take the maximum over the possibility of mirroring or not
and the possibility of early exercise or not. This, also, will in general
force us to use numerical methods like for example a binomial tree approach.

\bigskip

An important fact according to our results is that no matter what the market
conditions are, all mirror options with monotonic increasing payoffs or all
mirror options with monotonic decreasing payoffs will have the same initial
value as their standard counterparts. This is a really surprising result
since in this case probably nobody would buy the equal value standard
options! Moreover when we restrict the holder to make a maximum number of
mirrorings $M$, we have obtained that at the beginning of the contract the
value of any mirror option with monotonic payoff and $M\geq 1$ does not
depend on $M.$ Again this seems to be counterintuitive since we expect
holders to prefer mirror options with more allowed mirrorings than with
lesser ones. Even though these results are correct in our simplified market
model we expect corrections to appear in more involved models.

In particular inefficient market models\cite
{Fama,Wang,Serva,Zhang,Bouchaud,Matacz,Masoliver}, where we have
correlations between assets returns at different times, may lead to
significant deviations from our results depending on the degree of
inefficiency in the model. This can be expected on the intuitive basis that
partial knowledge of future returns can be used by the holder to improve his
performance with adequate mirrorings.

If inefficiencies can bring about corrections to our formulae the
incompleteness of the market \cite{Hofmann,ElKaroui,Neuberger,Aurell} is not
less important (at least equally important as it is in the valuation of
standard options). Incompleteness in the market can be generated by several
reasons, for example it can appear in mixed jump-diffusion models \cite
{Merton,Scott1,Aase,Kou}, stochastic volatility and term structure models 
\cite{HullWhite,Scott2,Stein,Heston,Ghysels,Derman,HandN}, etc. In any case
both from the point of view of future implementations of mirror options as
real OTC products or from the theoretical point of view it is also
interesting to consider the modifications these models can produce in our
valuation formulae.

Finally we cannot overview deviations appearing in our results in the
presence of transaction costs \cite{Leland,Hodges,Dewynne}. In particular we
have to remember that even though we have derived our formulae in the
absence of such costs it is just in the real world {\em with transaction
costs} where mirror options are attractive! Therefore we have to consider
that absence just as a simplification to the valuation procedure, in the
same way as with standard options.

Here it is worth to note the following: when analyzing the hedging of mirror
options in section \ref{hedging} we have realized that each mirroring made
by the holder implies a change of sign in the $\Delta $ of the hedger. This
is in fact intuitive since mirror options can be thought as a way to
transfer transactions costs from the holder to the issuer. However, since
the issuer deals in general with large portfolios where cancellations
between different positions occur, we are in fact effectively transferring
only a fraction of those costs. In general such fraction will be lower the
bigger the portfolio of mirror and standard options the issuer has. However
as we warned in section \ref{hedging} in situations of large illiquidity
(like those appearing in crashes or rallies) collective behavior of holders
of mirror option can lead to large changes in delta that can not be easily
canceled out. Because of that we have considered in section \ref{finiteM}
the valuation of mirror options when only a finite number $M$ of mirrorings
is allowed. Then we can take the value of a real mirror option to be the one
of section \ref{finiteM} (or the numerical result if the payoff is not
monotonic) plus a correction depending on $M$ and other factors
incorporating the inefficiency, incompleteness and transactions costs of the
market. However, as it well known, when considering transaction costs it is 
{\em the hedger's whole portfolio of options} (mirror and standard options)
what should be valued \cite{Dewynne}. What it is not well known, to our
knowledge, is how to ascribe a fraction of such value to each component of
the portfolio. This is necessary if we want to obtain a realistic correction
to our results due to transaction costs.

\bigskip 

Considering the above remarks we can fairly say that a lot of work can still
be done in order to have practical implementations of mirror options.
However, because of their attractive characteristics we believe that
innovative financial institutions can make real OTC mirror options available
in the near future.

\section{Acknowledgments}

J.M. would like to thank J. Masdemont, A. Gisbert, J. Masoliver and J.
Perell\'{o} for their useful comments. The author also acknowledges a
fellowship from Generalitat de Catalunya, grant 1998FI-00614.

\end{document}